\documentclass[sigconf]{acmart}

\copyrightyear{2024} 
\acmYear{2024} 
\setcopyright{acmlicensed}\acmConference[RecSys '24]{18th ACM Conference on Recommender Systems}{October 14--18, 2024}{Bari, Italy}
\acmBooktitle{18th ACM Conference on Recommender Systems (RecSys '24), October 14--18, 2024, Bari, Italy}
\acmDOI{10.1145/3640457.3688136}
\acmISBN{979-8-4007-0505-2/24/10}

\usepackage{enumitem}
\usepackage{multirow}
\usepackage{amsmath}
\usepackage{mathrsfs}
\usepackage{graphicx}
\usepackage[T1]{fontenc}
\usepackage{aecompl}
\usepackage{subfigure}

\AtBeginDocument{%
  }

\begin{document}
\title{AIE: Auction Information Enhanced Framework for CTR Prediction in Online Advertising}

\author{Yang Yang}
\authornote{Co-first authors with equal contributions. }
\email{yangyang590@huawei.com}
\affiliation{%
  \institution{Huawei Noah's Ark Lab}
  \country{China}
}
\author{Bo Chen}
 \authornotemark[1]
\email{chenbo116@huawei.com}
\affiliation{%
  \institution{Huawei Noah's Ark Lab}
  \country{China}
}

\author{Chenxu Zhu}
\email{zhuchenxu1@huawei.com}
\affiliation{%
  \institution{Huawei Noah's Ark Lab}
  \country{China}
}

\author{Menghui Zhu}
\email{zhumenghui1@huawei.com}
\affiliation{%
  \institution{Huawei Noah’s Ark Lab}
 \country{China}
}

\author{Xinyi Dai}
\email{daixinyi5@huawei.com}
\affiliation{%
  \institution{Huawei Noah’s Ark Lab}
 \country{China}
}

\author{Huifeng Guo}
\email{huifeng.guo@huawei.com}
\affiliation{%
  \institution{Huawei Noah’s Ark Lab}
 \country{China}
}

\author{Muyu Zhang}
\email{zhangmuyu@huawei.com}
\affiliation{%
  \institution{Huawei Technologies Co Ltd}
 \country{China}
}

\author{Zhenhua Dong}
\email{dongzhenhua@huawei.com}
\affiliation{%
  \institution{Huawei Noah’s Ark Lab}
 \country{China}
}

\author{Ruiming Tang}
\authornote{Corresponding authors.}
\email{tangruiming@huawei.com}
\affiliation{%
  \institution{Huawei Noah Ark's Lab}
 \country{China}
}

\renewcommand{\shortauthors}{Yang Yang et al.}

\begin{abstract}
Click-Through Rate (CTR) prediction is a fundamental technique for online advertising recommendation and the complex online competitive auction process also brings many difficulties to CTR optimization. Recent studies have shown that introducing posterior auction information contributes to the performance of CTR prediction. However, existing work doesn't fully capitalize on the benefits of auction information and overlooks the data bias brought by the auction, leading to biased and suboptimal results.
To address these limitations, we propose \textbf{A}uction \textbf{I}nformation \textbf{E}nhanced Framework (AIE) for CTR prediction in online advertising,
which delves into the problem of insufficient utilization of auction signals and first reveals the auction bias. 
Specifically, AIE introduces two pluggable modules, namely Adaptive Market-price Auxiliary Module (AM2) and Bid Calibration Module (BCM), which work collaboratively to excavate the posterior auction signals better and enhance the performance of CTR prediction. 
Furthermore, the two proposed modules are lightweight, model-agnostic, and friendly to inference latency. Extensive experiments are conducted on a public dataset and an industrial dataset to demonstrate the effectiveness and compatibility of AIE.
Besides, a one-month online A/B test in a large-scale advertising platform shows that AIE improves the base model by 5.76\% and 2.44\% in terms of eCPM and CTR, respectively. 

\end{abstract}

\ccsdesc[500]{Information systems~Retrieval models and ranking}

\keywords{Online Advertising, Click-through Rate Prediction, Posterior Feature Modeling}


\maketitle

\section{Introduction}

Click-Through Rate (CTR) prediction~\cite{richardson2007predicting, yang2022click} holds a vital place in online advertising systems since CTR prediction performance directly influences the overall satisfaction of the users and the revenue generated by companies. The CTR is the probability that a user clicks the ad when it is shown. Advertisers submit cost-per-click (CPC) bids~\cite{hu2016incentive, zhu2017optimized} to show how much they are willing to pay if a user clicks. The common practice for advertisement platforms is combining predicted click-through rates (pCTR) and CPC bid to calculate effective 
cost per mille (eCPM) for rank to maximize the expected revenue \cite{fridgeirsdottir2018cost,lyu2023pairwise}. In this process, CTR prediction is very important as it directly influences the auction's outcomes.

However, the complicated auction environment in online advertising systems poses great challenges to CTR prediction tasks. 

Firstly, the ranking criteria eCPM is determined by pCTR and CPC bid~\cite{hummel2017loss, zhu2017optimized}, which means CPC bid will also directly determine the final ranking results. The CPC bid changes over time and it influences the ranking results and user clicks. However, when we model CTR, this information is not explicitly considered, which leads to a biased estimation. Secondly, the auction environment is highly dynamic with many third-party demand-side platforms (DSPs) participating in the auction competition, causing fluctuations and instability in advertisement auction results. 

Typically, CTR prediction models highly rely on the offline training data collected and lack the ability to perceive bidding information and auction environment. Therefore, it is common in real-world industrial applications that a CTR model shows high accuracy offline but exhibits no significant improvement online.

\begin{figure}[hbtp]
\centering
\subfigure[CTR-Market Price relationship]{
\begin{minipage}[b]{0.43\linewidth}
\includegraphics[width=1\linewidth]{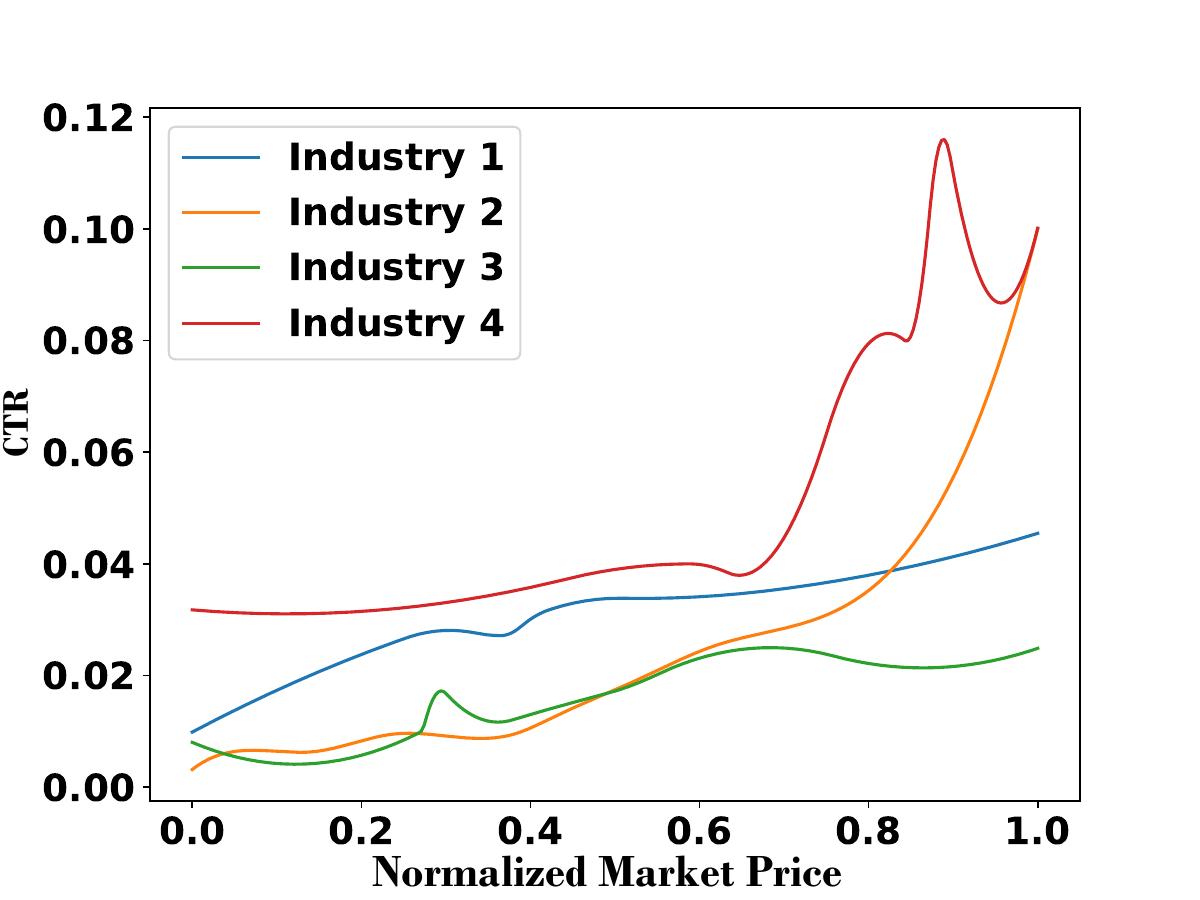}
\end{minipage}
}
\subfigure[Market Price Distribution]{
\begin{minipage}[b]{0.43\linewidth}
\includegraphics[width=1\linewidth]{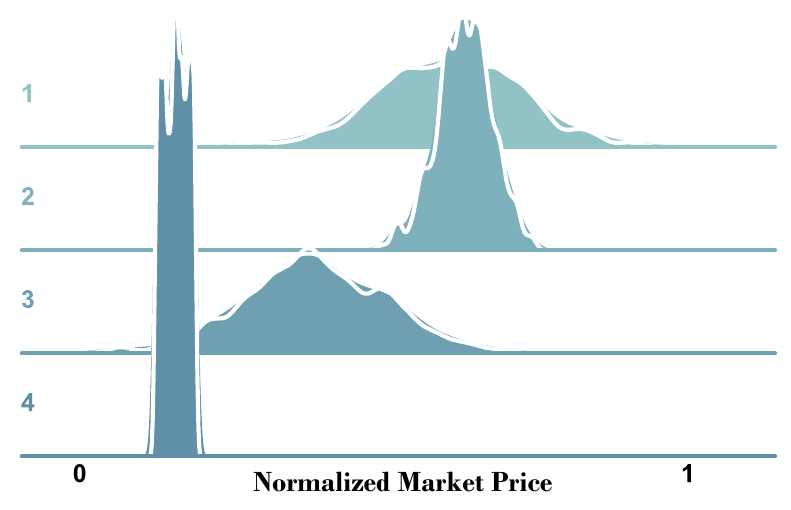}
\end{minipage}
}
\caption{(a) CTRs against different market prices under four industries in an industrial dataset. (b) Market price distribution under four industries in an industrial dataset.}
\label{fig:dist}
\end{figure}

To model the highly dynamic auction environment and mitigate the offline-online inconsistency, we need to take the auction environment and the bidding information into consideration in the phase of CTR prediction. CTR prediction and the market price modeling tasks have strong inherent interdependence~\cite{yang2021multi}. Figure 1 (a) shows the CTR is positively correlated with the market price under four industries in our industrial dataset. Some work\cite{yang2021multi} considers introducing market price in the CTR prediction phase but lacks fine-grained modeling to capture the variance of market price distributions under different scenarios, leading to sub-optimal results. The distribution of market prices varies significantly across different scenarios as shown in Figure 1 (b). We depict the market price distribution for the four selected industries in a real-world advertising platform, where the vertical axis divides scenarios and the horizontal axis shows the market prices' relative values. Therefore, how to exploit the auction signals including market price and features that distinguish the market environment to improve CTR prediction is a vital issue, which we call it \textbf{auction signals utilization problem}.

\begin{figure}[h]
    \vspace{-2mm}
    \centering
    \includegraphics[width=1.0\linewidth]{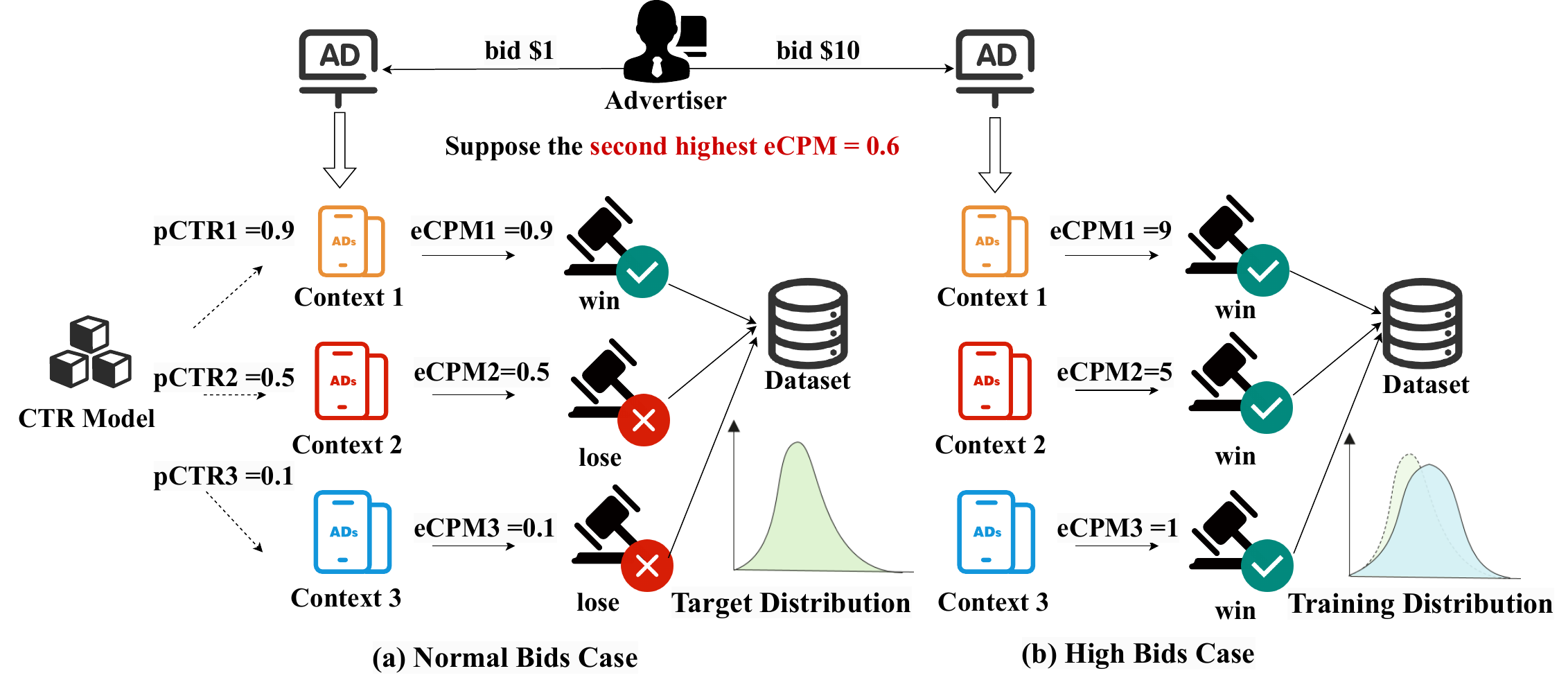}
    \caption{Auction Bias Illustration. When the advertiser gives a normal bid for an ad as Figure (a) shows, the ad only wins the auction in Context 1 due to the high click relevance. When the advertiser gives a high bid for the same ad as Figure (b) shows, the ad wins more auctions in all three Contexts though the click relevance is low for Context 2 and 3. Therefore, the training data distribution is different from the original target distribution, leading to data bias, which is called auction bias due to high bidding.}
    \label{fig:compare}
    \vspace{-1mm}
\end{figure}

Recommender systems are always subject to various biases, including popularity     bias~\cite{zhu2021popularity,zheng2021disentangling}, position bias~\cite{vardasbi2020cascade,wang2016learning}, selection bias~\cite{wang2021combating,wang2018deconfounded} and so on. In the context of advertisement recommendations, CTR prediction is facing a certain bias brought by the auction. Suppose some advertisers bid very high for their ads, which causes these ads can win the auction more easily even if the respective click relevance of them are lower than other ads. Although these ads gained more exposure, they are recommended to less relevant users who are not inclined to click. The over-exposure leads to a lower CTR for those ads with high bids.
These samples flow into our training data, causing the distribution of CTR of the ads with high bids to be shifted and biased. We call this kind of bias in training data as \textbf{auction bias}. Take Figure 2 as an example. Suppose we have a CTR model that can give basically accurate CTR predictions, it outputs three pCTRs 0.9, 0.5, and 0.1 under three different contexts for an advertisement. Here 
$eCPM=bid*pCTR$
and suppose the second highest eCPM from other competing DSPs for all contexts is 0.6. When the advertiser gives a normal bid as \$1 for this ad, it can win the auction only under context 1 as illustrated in Figure 2 (a). However, when the advertiser gives a high bid as \$10 for this ad, it can win under all three contexts and thus gain some unpreferable exposure since the click relevance for context 2 and context 3 are relatively low as Figure 2 (b) shows. This leads to the CTR of this ad in our collected data being lower than the unbiased situation as only exposure data can be collected in advertising recommendations. It is worth noting that the phenomenon of advertisers increasing their bids to gain exposure is widespread. Therefore, auction bias exists widely in the training data in the advertisement recommendation scenario, and dealing with it is meaningful and significant.

To address the auction signals utilization problem and the auction bias mentioned above, we propose a novel framework named Auction Information Enhanced Framework for CTR prediction in online advertising (AIE)  which is composed of two modules Adaptive Market-price Auxilary Module (AM2) and Bid Calibration Module (BCM). AM2 constructs an auxiliary task to make use of the market price with a dynamic network to capture the variance of market price across different scenarios. BCM alleviates the auction bias in our data by approaching the target distribution utilizing bid information reasonably.
Our paper contributes to the literature in the following ways:

\begin{itemize}
    \item We pay attention to the challenges brought by the auction environment in online advertising systems and first reveal auction bias and its influence on CTR prediction. 

    \item We propose a novel framework AIE to take full advantage of the posterior auction signals and alleviate the auction bias. AIE can improve the performance of CTR prediction models with two lightweight and model-agnostic modules.

    \item  Comprehensive experiments on a public dataset and an industrial dataset validate the superiority of AIE. Results on a large-scale online advertising system further confirm the effectiveness and applicability of AIE.
\end{itemize}

\section{Method}
In this section, we first describe the problem definition of the CTR prediction considering auction information under the advertisement recommendation scenario. Then we provide an overview of AIE, detail its key components and give a discussion about some interesting findings.

\subsection{Problem Formulation}
Considering a training dataset $\mathcal{D}=\left\{\left(\boldsymbol{x}_j, y_j\right)\right\}_{j=1}^{|\mathcal{D}|}$ with $|\mathcal{D}|$ samples, where  $\boldsymbol{x}_j=\{ c_1,...c_i,...c_I \}$ and $y_j$ represent the \textit{common features} for CTR prediction and binary click label of the $j_{th}$ sample, respectively. The task of CTR prediction in the common online advertisement is to build a prediction model to estimate the probability of a user clicking a specific ad in a given context, which can be formulated as $\hat{y} = CTR\_model(\boldsymbol{x})$ where $\boldsymbol{x}$ is the common feature about user, ad and context to predict relevance.

However, CTR models normally ignore any posterior auction information and give predictions merely based on common features, causing severe problems as we can see in Figure 2. Hence we need to consider this posterior information in the training phase. Features $\boldsymbol{a} $ represents the \textit{posterior auction information} including market price, CPC bid and so on. Features $\boldsymbol{s}$ represents scenario-related features that indicate auction environments. Auction information can only be used in offline training because they are posterior feature for CTR prediction models. Combined with the common feature $\boldsymbol{x}$, the task we defined here in advertisement recommendation can be formulated as $\hat{y} = CTR\_model(\boldsymbol{x}, \boldsymbol{a}, \boldsymbol{s})$. Our goal is to enhance the CTR prediction model's performance by taking advantage of the extra auction information.

\subsection{Overview}
In this section, we present the overview architecture of AIE as illustrated in Figure 3. Since the auction information is posterior for the CTR prediction model, we only utilize it in the training phase and design two pluggable modules that do not take effect during the inference phase. 
The Adaptive Market-price Auxiliary Module (AM2) and the Bid Calibration Module (BCM) are deployed to leverage auction information to boost CTR prediction models' performance. The Adaptive Market-price Auxiliary Module(AM2) is deployed to let the model learn useful knowledge in auction information by a multi-objective structure.

\begin{figure*}[htbp]   
    \centering
    \vspace{-2pt}
\includegraphics[width=0.9\linewidth,scale=1.00]{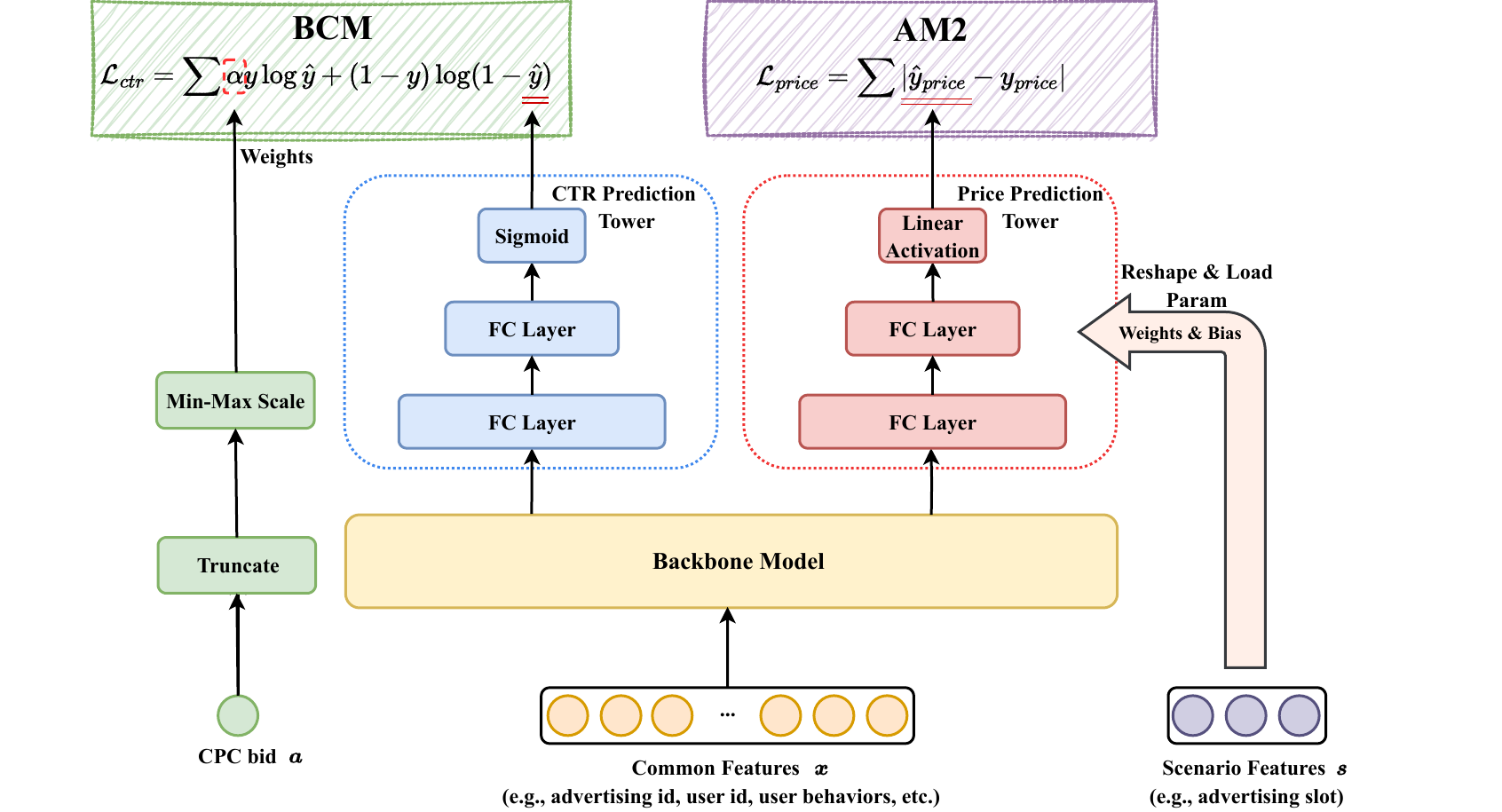}
    \caption{Overall framework of AIE, which consists of two key modules: AM2 and BCM. AM2 uses market price and scenario features to construct an auxiliary task. BCM uses bid to impact the cross-entropy loss.}
    \label{figure/1.png}
    \vspace{5pt}
\end{figure*}

\subsection{Adaptive Market-price Auxiliary Module}

The always-changing auction environment and the fierce competition in the market under advertisement recommendations pose great challenges to CTR prediction models. The rank eCPM can be influenced greatly by auction-related factors thus leading to results that fall short of expectations.
A key problem we face in advertisement recommendation is how to utilize the various auction information to empower the CTR prediction model. Hereby, we propose AM2 to model the auction information at a fine-grained level. 

We observe the phenomenon that market price and the actual CTR are to some extent positively correlated in our industrial application as Fig 1(a) shows. By designing an auxiliary task to fit the market price as complementary to the main CTR prediction task based on a shared bottom structure, 
useful information can be learned through the shared embedding and the backbone model. Moreover, to capture the variance of market price distribution, the price prediction tower's weights and biases are generated by a dynamic weight network fed by scenario features that can differentiate auction environments \cite{yang2022adasparse, zhang2022leaving, gao2023scenario}. Given the input of the price prediction tower $\mathbf{h}^{(0)}$, which is the last layer representation of the backbone model, the price prediction tower's MLP can be formulated as: 
\begin{equation}
    \mathbf{h}^{(k)} = \sigma(\mathbf{W^{(k-1)}} \mathbf{h}^{(k-1)} + \mathbf{b^{(k-1)}}),\quad k\in[1,\cdots, N]~,
\end{equation}
\begin{equation}
\mathbf{W^{(k)}}, \mathbf{b^{(k)}} = Reshape(Split(\mathbf{E}^s))~,
    \end{equation}
where $\sigma$ is the activation function, $N$ is the depth of the MLP, $\mathbf{E}^s$ is the designated scenario embedding. The output of the last layer of the MLP $\mathbf{h}^{(N)}$ is $\mathbf{\hat{y}_{price}}$, which is the estimated value of the market price. In particular, the weight and bias parameters $\mathbf{W^{(k)}}$ and $\mathbf{b^{(k)}}$ are generated by the designated scenario embedding by reshaping and splitting. The number of layers and 
hidden units of each layer of the MLP are hyperparameters that need to be tuned. In our experience, a lightweight MLP is enough for the price prediction tower and the CTR prediction tower to gain a good overall performance. 
A visual illustration of this process is shown in the right part of Figure 3.

After we get the estimated value of the market price, we compare it with the real value of the market price and calculate the regression loss, which can be expressed as:
\begin{equation}
    \mathcal{L}_{price}=\sum{ |\hat{y}_{price}-{y}_{price}|}.
\end{equation}
Here, we use the MAE regression loss because it doesn't rely on prior knowledge like statistical bucketing of market prices and performs better than classification loss.
Then, AIE can be jointly trained by optimizing the weighted sum of the ctr prediction loss and the price prediction loss. The final loss can be demonstrated as:
 \begin{equation}
    \mathcal{L}=\mathcal{L}_{ctr}+w*\mathcal{L}_{price},
\end{equation}
where $w$ is the hyperparameter that 
controls the importance of the price prediction auxiliary task. Because our goal is to enhance the CTR prediction's performance, the price prediction task's performance doesn't matter. So we tune the $w$ to ensure the ctr prediction task performs the best. 

\subsection{Bid Calibration Module}


CTR prediction is one of the core algorithms in computational advertising. In the CPC billing model, the mechanism design can simply sort the advertisements by effective eCPM to maximize advertising revenue. CPC bids directly affect the final rank eCPM of the advertisement, thus affecting the advertisement's exposure. As illustrated in Figure 2, high bids given by advertisers can lead to more exposures, some of these are low-quality exposures and 
resulting in a lower CTR. 

Intuitively, we can reweight the training sample to calibrate the biased training distribution. Specifically, we assign higher weights for positive samples with higher CPC bids. 
Here we only reweight the positive samples because they can influence the positive ratios in training data and they are worth paying more attention to since users click on them.
In this way, we raise the CTR in the training data for those samples with high bids to approach the target distribution. 
There are always bids with very large values in some extreme cases and we need to clip the origin value to avoid the effects of extreme values on our training loss.
Then the clipped values are transformed by Min-Max scaling \cite{ahsan2021effect} to a specific range as the final weighting factor. To control the prediction bias, the overall expectation of the positive ratio after reweighting remains the same as the original, which means some positive samples' weights with low bids would be decreased. The whole process can be divided into two steps. The first step, which is truncate can be formulated as:
 \begin{equation}
    \mathcal{C'}= clip(\mathcal{C},{min, max}),
\end{equation}
where $\mathcal{C}$ represents the original CPC bid value at sample-level, $min$, $max$ means the statistical minimum value and maximum value of $\mathcal{C}$ after removing the outliers and $\mathcal{C'}$ represents the clipped CPC bid value. Afterward, $\mathcal{C'}$ needs to be transformed to a designated range, which can be expressed as:
 \begin{equation}
    \alpha= a + \frac{(b-a)}{(max-min)} * (\mathcal{C'}-min),
\end{equation}
where $a$, $b$ means the lower and upper bound that specify the interval of the transformation and $\alpha$ means the reweighting factor we get. It's worth noting that in real industry applications, the bid's distribution varies under different scenes. In that case, we need to calculate $\alpha$ for each scene and perform fine-grained weighting. Then we take advantage of this additional information to reweight the positive samples, which can be reflected in the ctr loss as:
 \begin{equation}
\mathcal{L}_{ctr}=\sum{\alpha y\log \hat{y}+(1-y)\log(1-\hat{y})},
\end{equation} 
where $\alpha$ here is the final weighting term for each sample. By taking the CPC bids as auxiliary information into account, we calibrate the training CTR distribution towards the target distribution, thus making the CTR prediction more accurate. 
Our approach is more suitable for scenarios where advertisers do not adjust their bidding strategies frequently. In our industrial practice, advertisers adjust their bids daily, and our model can be updated on an hourly basis or even faster to make sure BCM captures the real-time auction information.
More interesting perspectives on this method will be presented and discussed in Section 2.5.

\subsection{Discussion}
Firstly, we would like to discuss the BCM's connections and differences with the previous debias methods like inverse-propensity-scoring (IPS). IPS is a practical debias method for industry products, which can be regarded as a specific case of reweight learning. In our case, the propensity score can be defined as:
 \begin{equation}
    p_i = \frac{P(x_i,y_i)}{Q(x_i,y_i)},
\end{equation}
where $P(x,y)$ is the training distribution, $Q(x,y)$ is the target distribution without the impact of the bidding-related factors. However, in advertisement recommendation, it's impossible to rule out the impact of bidding factors and observe the unbiased distribution. So here we deployed BCM to calibrate the training distribution as mentioned in Section 2.4. It's similar to the IPS format and the optimal function can be expressed as:
 \begin{equation}
    \hat\theta=argmin_{\theta\in\Theta}\Sigma\frac{l(z_i,\theta)}{p_i}+\lambda R(\theta)
\end{equation}
where $z_i=(x_i,y_i)$ is an observed sample drawn from training distribution, $\theta\in\Theta$ is a model parameter, $l(z_i,\theta)$ is a loss function, $R(\theta)$ is a regularizer. $p_i$ here is approximated by the inverse of the weighting term $\alpha$ for each sample.

Secondly, we present another perspective on what BCM is doing. BCM raises the CTR for those positive samples with high bids. If we reckon that items with higher bids have higher value in the online advertising system, pCTR for those high-value items would also be raised. When competing with other third-party DSPs regarding high-value traffic, the platform would gain an advantage and have a better chance of winning. In this way, the revenue of the platform will be increased.

\section{Experiments}
In this section, we present our experiments in detail, including experimental setup, model
comparison, and the corresponding analysis. We conduct experiments on a public dataset and an industrial dataset to investigate the following questions:
\begin{itemize}[leftmargin=*]
    \item \textbf{RQ1:} How much does our framework enhance the accuracy and revenue compared to competing methods?
    \item \textbf{RQ2:} Is the proposed framework suitable for SOTA backbone CTR models? 
    \item \textbf{RQ3:} How effective are the two proposed modules (i.e., AM2 and BCM) for improving the performance?
    \item \textbf{RQ4:} How are the AIE's training and inference efficiency?
\end{itemize}
\subsection{Experimental Setting}

\subsubsection{Datasets}

\begin{itemize}
\item \textbf{iPinYou}\footnote{https://contest.ipinyou.com/}
The iPinYou dataset \cite{liao2014ipinyou} is a real-world
dataset for ad click logs over 10 days. After one-hot encoding,
we get a dataset containing 19.50M instances with 937.67K
input dimensions. We keep the original train/test splitting
scheme, where for each advertiser the last 3-day data are used
as the test dataset while the rest as the training dataset. We follow
the previous data processing \cite{qu2016product}. To the best of our knowledge, it is the most appropriate public dataset for AIE to valid its performance, which contains \textit{paying price} (the highest bid from competitors, also called market price and auction winning price), \textit{bidding price} (the bid price from iPinYou for this bid request) \cite{zhang2014real} information and enough other user, item features for CTR prediction.
\end{itemize}

\subsubsection{Evaluation Metrics}
Four evaluation metrics are tested in our experiments. The
two major metrics are:
\begin{itemize}
\item \textbf{AUC}\ Area under ROC curve is a widely used metric in evaluating classification problems. Specifically, a higher AUC value at the ``0.001'' level indicates significantly better performance~\cite{guo2017deepfm}.

\item \textbf{csAUC}\ CPM-sensitive AUC (csAUC) \cite{liu2019cpm} is a metric that takes bid into consideration in offline evaluation to better approach the online evaluation metric eCPM for advertising platforms. If a high-level sample $x_h$ and a low-level sample $x_l$ are randomly selected from dataset $\mathcal{D}$. Given our revenue of $(x_h, x_l)$ as follows:
 \begin{equation}
Rev(x_{h}, x_{l}) = \begin{cases}bid_{h} & pCTR_h*bid_h >= pCTR_l*bid_l\\T(x_l) &, otherwise\end{cases}
\end{equation}
 \begin{equation}
T(x_i)=\begin{cases}0 & if\ x_i\ is\ a\ negative\ sample\\bid_i &otherwise\end{cases},
\end{equation}
csAUC of dataset $\mathcal{D}$ is $\frac{\sum_{(x_h, x_l)\in D} Rev(x_h, x_l)}{\sum_{(x_h, x_l)\in D} bid_h}$.
 csAUC not only takes the ability of the model to distinguish between positive and negative samples into consideration but also evaluates if the model is able to rank the advertisement with a high value higher among the positive samples.
Following the method proposed in \cite{liu2019cpm}, we can calculate csAUC on large-scale data in real-world data applications. 

\item \textbf{Rev}\  We define a metric called Rev on iPinYou dataset to simulate the revenue that a prediction model would generate for each auction by ad slot granularity similar to the Rev defined in \cite{wu2018turning}. Because we only have exposure data, from which we need to simulate bidding and exposure process and then count revenue based on that. Following the previously proposed method \cite{ou2023deep}, we group the testset logs by the following features: \textit{weekday}, \textit{hour}, \textit{slotid} to simulate an auction. Then we rank each group's samples by the simulated eCPM ($eCPM = pCTR * bid$). In the single-slot advertisement scene, only the advertisement with the highest eCPM wins the auction. If that top 1 ad was clicked in our test set, we would calculate the \textit{paying price}, which can be formulated as:
 \begin{equation}
   i = \arg\max_g{pCTR_g * bid_g} \quad g\in G,
\end{equation}
 \begin{equation}
   Rev_g = \begin{cases}payprice_{i} & click=1\\0 & click=0\end{cases},
\end{equation}
 \begin{equation}
   Rev = \sum_{{\forall}g\in G}Rev_g,
\end{equation}
where $g$ means one simulated auction across all groups $G$, $i$ means the item with the highest eCPM within that group $g$, $Rev_g$ means the simulated revenue for group $g$ for the top 1 ad recommendation. $Rev$ is the sum of $Rev_g$ across all groups in testset. This metric is a good representation of the platform's profit.
\item \textbf{Rev NDCG}\ Following the principles of the NDCG \cite{krichene2020sampled}, we proposed Rev NDCG. Based on the Rev obtained, we divide it by the Maximum Possible Revenue gained by the platform to get Rev NDCG \cite{wu2018turning}. The Maximum Possible Revenue is defined as the sum of \textit{paying price} of the clicked item with the highest \textit{paying price} within each group, which can be formulated as:
 \begin{equation}
   Rev_{g_{max}} = \max (paying price_{i}), \quad    i\in ClickedItems
\end{equation}
 \begin{equation}
   Rev_{max} = \sum_{{\forall}g\in G}Rev_{g_{max}},
\end{equation}
 \begin{equation}
   Rev NDCG = \frac{Rev}{Rev_{max}},
\end{equation}
where $Rev_{g_{max}}$ is the Maximum Possible Revenue gained by the platform for one auction $g$, $Rev_{max}$ is the Maximum Possible Revenue gained by the platform among all auction groups.

\end{itemize}

\begin{table*}[h]
	\centering
	\vspace{-1.0em}
	\caption{Overall performance comparison on different backbones on iPinYou dataset. Boldface denotes the highest score and * represents significance level $p$-value < 0.05.}
	\vspace{-0.6em}
	\label{tab:baseline}
	\resizebox{0.9\textwidth}{!}{
	\begin{tabular}{c|c|ccccccc|c}
		\toprule
		 Metric & Model & DNN & DCN & DeepFM & AutoInt &  FiBiNET &  DCNV2& DFFM & RelaImpr. (Avg)\\
		\midrule
		 \multicolumn{1}{c|}{\multirow{3}{*}{AUC}} & Baseline &0.7755 & 0.7770 & 0.7757 &0.7738 &0.7763&0.7766&0.7766&- \\
		&   MTAE &0.7780 & 0.7769 & 0.7759 &0.7772 &0.7763&0.7776&0.7757&0.32\% \\   &AIE &\textbf{0.7792* } & \textbf{0.7797* } & \textbf{0.7780* } & \textbf{0.7793* } & \textbf{0.7785* } & \textbf{0.7781* } & \textbf{0.7782* } &\textbf{1.01\%} \\
		\cmidrule{1-10}
		 \multicolumn{1}{c|}{\multirow{3}{*}{csAUC}} & Baseline &0.7762 & 0.7777 & 0.7763 &0.7743 &0.7768&0.7771&0.7775 &-\\
		  & MTAE &0.7792 & 0.7770 & 0.7768 &0.7777 &0.7768&0.7781&0.7761&0.30\% \\  &AIE &\textbf{0.7798* } & \textbf{0.7804* } & \textbf{0.7786* } & \textbf{0.7798* } & \textbf{0.7792* } & \textbf{0.7787* } & \textbf{0.7787* } &\textbf{1.00\% } \\
		\cmidrule{1-10}
		 \multicolumn{1}{c|}{\multirow{3}{*}{Rev}} & Baseline  & 26.489 &27.317 & 27.640&26.778 &26.544 & 27.681& 26.440 &-\\
		  & MTAE  & 27.947 &27.591 &26.147 &27.641 &27.376 &27.386 &27.476 &1.47\% \\
		  & AIE  & \textbf{28.428* } &\textbf{28.355* } &\textbf{27.652* } &\textbf{28.536* } &\textbf{27.808* } &\textbf{27.954* } &\textbf{27.540* } &\textbf{3.95\%}\\
		\cmidrule{1-10}
		 \multicolumn{1}{c|}{\multirow{3}{*}{Rev NDCG}} & Baseline  & 0.1249 &0.1289 & 0.1304&0.1263 &0.1252 & 0.1306& 0.1247&- \\
		  & MTAE  & 0.1318 &0.1301 &0.1233 &0.1304 &0.1291 &0.1292 &0.1296&1.46\% \\
		  & AIE  & \textbf{0.1341* } &\textbf{0.1338* } &\textbf{0.1305* } &\textbf{0.1346* } &\textbf{0.1312* } &\textbf{0.1319* } &\textbf{0.1299* } &\textbf{3.97\%}\\
		\bottomrule
	\end{tabular}
}

\vspace{1em}
\end{table*}

\subsubsection{Competing Models}

To verify the effectiveness of the proposed approach, we compare AIE with the following baselines.
The backbone CTR prediction models we selected include DNN \cite{taud2018multilayer}, DCN \cite{wang2017deep}, DeepFM \cite{guo2017deepfm}, AutoInt \cite{song2019autoint}, FiBiNET \cite{huang2019FiBiNET}, DCNv2 \cite{huang2019FiBiNET}, DFFM \cite{guo2023dffm}, which includes the classical CTR prediction models that captures different orders of feature interactions like DCN, DeepFM and advanced multi-scenario model like DFFM. We also include Multi-task Advertising Estimator (MTAE) \cite{yang2021multi} as a competing model since it introduces market price information to jointly optimize CTR prediction.

\subsubsection{Implementation Details}

We use Adam \cite{kingma2014adam} optimizer to optimize different models.
For fair comparison, we fix the embedding size as 8, the batch size as 2000 for all models. The hyperparameters of the deep layer are tuned for each model and ensure that the capacity is about the same.
For CTR prediction tower and price prediction tower's micro MLP, the number of layers is searched from 1 to 3, and neurons at each layer from \{16, 32, 64\}. The hyperparameter $w$ that controls the importance of the auxiliary price prediction loss is searched from 1e-5 to 1e-3. The lower bound $a$ for BCM is chosen from  0.5 to 1 and the upper bound is chosen from 1 to 2.
The learning rate is searched from \{1e-3,1e-4,1e-5,1e-6\}
and $L_{2}$ regularization coefficient from \{1e-4, 1e-5, 1e-6, 1e-7\}. Besides, we run each experiment 5 times with the optimal parameters searched and report the average performance.
Besides, RelaImpr~\cite{yan2014coupled} is applied to measure the relative improvement between the measured model and base model in terms of AUC and csAUC :
\begin{equation}
\text { RelaImpr }=\left(\frac{A U C(\text{Measured Model})-0.5}{A U C(\text {Base Model})-0.5}-1\right) \times 100 \%.
\end{equation}
As for the Revenue metrics, the RelaImpr is defined as :
\begin{equation}
\text { RelaImpr }=\left(\frac{Rev(\text{Measured Model})}{Rev(\text {Base Model})}-1\right) \times 100 \%.
\end{equation}
The code is available based on MindSpore \footnote{https://github.com/mindspore-lab/models/tree/master/research/huawei-noah/AIE}. We thank MindSpore ~\cite{mindspore} for the partial support of this work.

\subsection{Overall Performance (RQ1\&RQ2)}
This subsection gives an overall comparison between AIE and different baselines from the aspect of CTR prediction and revenue evaluation, whose results are depicted in Table 1. From this we can conclude that: 

\begin{itemize}[leftmargin=*]
    \item \textbf{CTR prediction} \quad From Table~\ref{tab:baseline},it can be observed that AIE
achieves significant improvement over MTAE and different baseline models of AUC. AIE and MTAE can both be adapted to any baseline model.  MTAE performs equal or better compared to the baseline models, which proves that the 
modeling of auction information is necessary and gainful. AIE  consistently yields better performance in terms of AUC based on all backbone models. Specifically, AIE enhances the AUC of AutoInt the most with a RelaImpr of 2.01\% and gains 1.01\% RelaImpr on average for all base models. Compared to MTAE, the average RealImpr on all base models increases from 0.32\% to 1.01\%, which is significant. This is due to our more fine-grained modeling of auction information.

    \item \textbf{Revenue Evaluation} \quad We can observe from Table~\ref{tab:baseline} that the trend of csAUC is basically the same as AUC. MTAE beats the baseline in most cases, proving the effectiveness of setting of adding an auxiliary task to predict market price. However, MTAE's improvement is not stable on some baselines like DCN, which is probably because it adopts a multi-classification loss whose effect
    depends on the precise grasping of the way market prices are categorized. AIE adopts regression loss which can avoid bucketing the market price to multi-class by manual experience. AIE gains 1.00\% RealImpr on csAUC since it achieves more comprehensive modeling of auction information.
   The presented Rev is divided by 1000 \cite{wu2018turning}. AIE performs the best for Rev and Rev NDCG on all backlines selected. The average RelaImprs of AIE on Rev and Rev NDCG are 3.95\% and 3.97\%, demonstrating a strong ability to optimize revenue. 
   \item \textbf{Recommend Tendency Exploration} \quad To figure out what kind of ads AIE tends to recommend to users, we conduct an experiment. 
 As Table~\ref{tab:recommend item} shown, we divide the advertisements in testset into three groups ranked by \textit{bidding price}, where the top 1/3 are categorized as 'high', the lowest 1/3 as 'medium' and the remaining as 'low'. We choose DCN as the baseline and then
calculate the proportion of ads that are being recommended for baseline and baseline+AIE. The following observations can be made: (i) the proportion of ‘high’ and 'medium' group items recommended to users significantly increases on average by AIE;
(ii) in contrast, the proportion of ‘low’ group items
recommended decreases by AIE. The overall revenue gain is positive because 'medium' and 'high' items are recommended more frequently and bring much more value to the platform.
These results align with the intuition of the AIE, which pays more attention to the high-bid samples at the training phase, thereby recommending high-value items to more users.
Combined with the AUC improvement illustrated in the \textbf{CTR prediction}, we can conclude that AIE not only enhances the sorting ability of CTR prediction model but also enables the CTR prediction model to perceive high-value traffic and thus improve the revenue.

\begin{table}[h]
  \vspace{0.6em}
	\caption{Distribution of items recommended to users across different groups of items divided by value on iPinYou testset. The basline here is DCN.}
	\vspace{-0.6em}
	\label{tab:recommend item}
	\centering
	\resizebox{0.7\linewidth}{!}{
		\begin{tabular}{c|cccccc}
			\toprule
            \multicolumn{1}{c|}{\multirow{2}{*}{Model}} & \multicolumn{3}{c}{iPinYou} \\ 
            \cmidrule(lr){2-4}
              &  Low & Medium & High   \\
            \midrule
            Baseline(\%)  & 57.8  & 2.3 & 39.9  \\
            Baseline+AIE(\%) & 52.2 & 5.1  & 42.7    \\
            Gain(\%) & -5.6   & +3.2& +3.5     \\
			\bottomrule
		\end{tabular}
		}
		\vspace{-0.6em}
\end{table}

    \item \textbf{Compatibility Analysis (RQ2)} As Figure 1 shows, BCM and AM2 can be adapted to any CTR prediction models as plug-ins. 
We apply AIE to seven CTR prediction models, namely DNN \cite{taud2018multilayer}, DCN \cite{wang2017deep}, DeepFM \cite{guo2017deepfm}, AutoInt \cite{song2019autoint}, FiBiNET \cite{huang2019FiBiNET}, DCNv2 \cite{wang2021dcn}, DFFM \cite{guo2023dffm} on iPinYou dataset to test its transferability.
As shown in Table 1, employing AIE to utilize auction information to assist the CTR prediction achieves much better performance on these seven models on iPinYou dataset. Therefore, we can conclude that AIE module has a strong compatibility with any CTR prediction models. 
   
\end{itemize}

\subsection{Ablation Study (RQ3)}

\begin{figure}[htbp]   
    \centering
    \vspace{-15pt}
\includegraphics[width=\linewidth,scale=1.2]{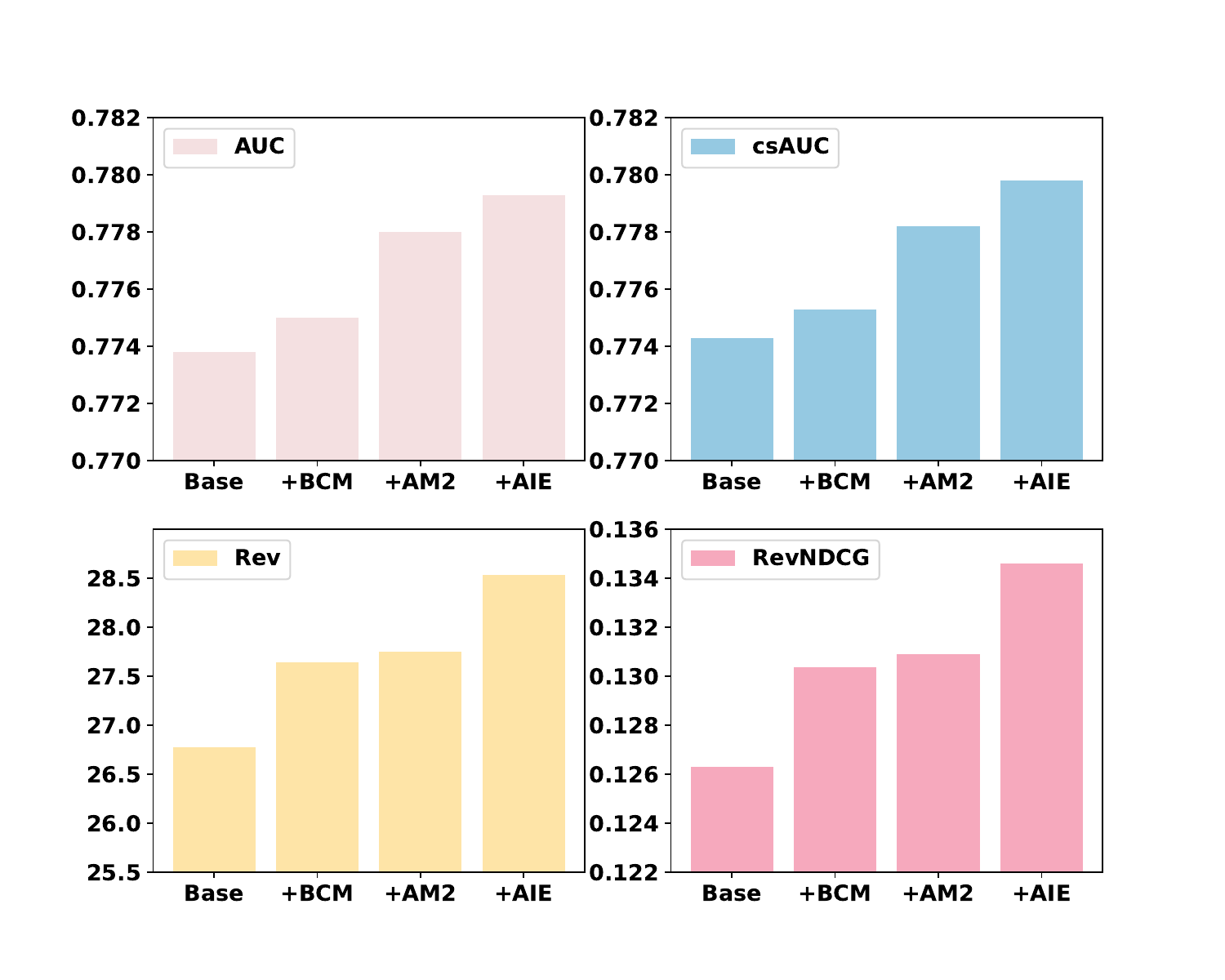}
    \caption{Ablation study about different modules of AIE in terms of four metrics. }
    \label{fig:ab1}
     \vspace{-10pt}
\end{figure}

We take AutoInt as the base model and verify the superiority of BCM and AM2 on this base model, respectively. 
The results are shown in Figure~\ref{fig:ab1}, from which we can draw the following conclusions.
\begin{itemize}
    \item Comparing BCM with the base model, we can observe that it can improve the revenue significantly and improve prediction accuracy slightly. This phenomenon is consistent with our design intuition of BCM, which leverages bids to alleviate the auction bias and improve the prediction accuracy. The offline metrics including AUC and csAUC are not improved significantly because the testing data is also biased and 
   hard to reflect the real effect.
    \item AM2 achieved significant improvement in sorting ability compared to the base model, including the ability to rank high-value positive samples higher. The revenue increase of AM2 is also considerable, manifesting that the fine-grained modeling of extra auction information is profitable for increasing revenue.
    \item BCM and AM2 both contribute substantially to the overall performance of AIE. 
    AIE achieves the best performance over these variants, confirming that leveraging posterior auction information in online advertising can boost the performance of CTR prediction model.
\end{itemize}

\subsection{Application in Industry System}

The industrial dataset is collected from a large-scale industrial Advertisement System, which samples from user behavior logs in eight consecutive days. We select the first seven days as the training set, sample part of instances as the validation set, and collect the last day as the testing set. This dataset contains item features
(e.g., creative ID, category), user features (e.g., user’s behavior history), context features(e.g., slot ID) and auction information(e.g., market price, bid). The deploy scenario contains hundreds of sites and mobile applications, where millions of daily active users interact with ads and tens of millions of user logs are generated every day.

\subsubsection{Offline Industrial Experiments}
Our framework is conducted on an offline industrial dataset, which is a large-scale dataset sampled from the click logs of the online advertisement platform and has more than 600 million impressions. We split them into train/valid/test sets by timestamp with a 6:1:1 proportion. Apart from common features $\boldsymbol{x}$ as defined in Section 2.1, we also collected auction information $\boldsymbol{a}$ including market price and CPC bid at sample level in our offline training data. 
In our scenario, the platform will bid and adjust bids on behalf of the advertiser based on pctr or pcvr, which leads to the fact that the bid can not be used during CTR prediction. The market price is only available after the auction stage, which is more posterior. Considering that, the bid and market price can not directly used as a feature for CTR prediction in most online ads recommendation system.
The scenario features $\boldsymbol{s}$ can be multiple features that reflect the auction environment like slot ID, app category,  hour of day and so on.


We use AUC and csAUC here to evaluate the offline performance of AIE. The baseline we compared with includes FiBiNET~\cite{huang2019FiBiNET}, DCN~\cite{wang2017deep}, EDCN~\cite{chen2021enhancing}, DFFM~\cite{guo2023dffm} and HierRec~\cite{gao2023scenario}.
The performance concerning revenue is evaluated on an online A/B test and will be presented later.

From Table~\ref{tab:industrial dataset result}, we summarize the observations from two dimensions.
From the horizontal dimension of Table~\ref{tab:industrial dataset result}, we can see that scenario-aware dynamic network is useful as DFFM and HierRec perform well.
From the vertical dimension of Table~\ref{tab:industrial dataset result}, the conclusion that AIE outperforms each baseline significantly can be easily drawn. The improvement in AUC shows AIE enhances the CTR model's sorting ability to distinguish the positive and negative samples. Moreover, AIE empowers the CTR model to better perceive high-value positive samples which can be reflected by the increase in csAUC.

\begin{table}[h]
	\centering
	\vspace{1.0em}
	\caption{Overall performance comparison on the industrial dataset.}
	\vspace{-0.6em}
	\label{tab:industrial dataset result}
	\resizebox{\linewidth}{!}{
	\begin{tabular}{c|c|ccccc}
		\toprule
		Metric & Model &  FiBiNET & DCN & EDCN & DFFM & HierRec  \\
		\midrule
		 \multicolumn{1}{c|}{\multirow{2}{*}{AUC}} & Baseline &0.8207 &0.8219 & 0.8226 & 0.8228 &0.8239 \\
	    &AIE &\textbf{0.8225}&\textbf{0.8231} & \textbf{0.8240} & \textbf{0.8239} & \textbf{0.8249}     \\
		\cmidrule{1-7}
		 \multicolumn{1}{c|}{\multirow{2}{*}{csAUC}} & Baseline &0.8296&0.8296 & 0.8299 & 0.8299 &0.8305  \\
          &AIE &\textbf{0.8323}&\textbf{0.8305} & \textbf{0.8310} & \textbf{0.8311} & \textbf{0.8315}    \\
		\bottomrule
	\end{tabular}
}
\vspace{-0.6em}
\end{table}

\subsubsection{Online Industrial A/B Test}
To evaluate the performance of our framework (i.e., AM2 and BCM) in the real industry application, we conduct an online A/B test in our online advertising platform for one month. AM2 was implemented in the first two weeks. Based on that, we added BCM and the complete framework AIE was deployed in the last two weeks.
The compared baseline is a highly optimized CTR model.  
Each model is trained over the latest exposure log, where an identical data process procedure is performed to ensure comparability. For online serving, all the three models (Base, AM2, AIE) are allocated 5\% of the overall traffic. Our models perform streaming incremental training hourly to capture real-time changes in auction information.

We compare the performance according to 
four metrics: RPM (Revenue Per Mille), eCPM (effective Cost Per Mille), CTR (Click Through Rate), and predicted bias, which are all widely used metrics for online advertising. Among them, CTR measures the relevance between the user and the ad.
RPM and eCPM are the core metrics for an advertising platform to measure revenue.
The predicted bias is calculated by $ \frac{(pCTR-CTR)}{CTR} \times 100 \% $, which 
needs to be kept within reasonable limits because excessive bias will infringe on advertisers' interests.

Table~\ref{tab:online} shows the results of the three models, among which AIE achieves 6.14\% improvements in RPM, 5.76\% improvements in eCPM and 2.44\% improvements in CTR while the predicted bias decreased by 5.38\%. 
The full volume of AIE went from 5\%, 10\%, 20\%, 50\% to 100\% of the flow, with each step being observed for 2-3 days and within each observation interval the enhancement of AIE is confident.  In Table 4, the confidence intervals for the improvement rate of AM2 and AIE regarding eCPM are [2.36\%, 5.14\%] and [4.75\%, 6.51\%]. Both the upper and lower bounds of the confidence interval are positive, indicating that the experimental results are statistically significant.

These results sufficiently validate that our framework can enhance the CTR prediction model's accuracy and revenue-optimization-oriented capabilities. Furthermore, the predicted bias of AIE decreased greatly compared to the base model, which proves the effectiveness of BCM in alleviating auction bias. 

The AM2 also outperforms the base model 1.83\%, 3.94\%, 1.21\% on RPM, eCPM, CTR respectively, proving the effectiveness of the elaborated way to use auction information.
After one month of evaluation, the AIE has become the main model in this scenario to carry all of the online traffic. 

\begin{table}[h]
	\centering
	\vspace{0.4em}
	\caption{Online A/B testing results of AM2 and BCM modules compared to the base model.}
	\vspace{-1em}
	\label{tab:online}
	\resizebox{0.9\linewidth}{!}{
	\begin{tabular}{c|cccc}
		\toprule
		 Model & RPM & eCPM & CTR & Bias \\
		\midrule
		 \multicolumn{1}{c|}{\multirow{1}{*}{Base+AM2}}  & +1.83\% &+3.94\% &+1.21\% &+2.58\%\\
	     \multicolumn{1}{c|}{\multirow{1}{*}{Base+AM2+BCM}}  &  +6.14\% & +5.76\% & +2.44\%  &-5.38\% \\

		\bottomrule
	\end{tabular}
}
\vspace{-0.6em}
\end{table}

\subsection{Efficiency Analysis (RQ4)}
In practical applications, the inference efficiency of the CTR prediction model is important since the recommender system has a high demand for real-time response. The training efficiency is also important because it affects how long it takes to update our model.
Therefore, to answer RQ4, this subsection presents a comparison of training and inference time between AIE and other baselines on the industrial dataset, whose results are summarized in Table~\ref{tab:eff}. 
Based on the results, it can be concluded that AIE's training time is essentially comparable to the baseline. Due to the lightweight design of the CTR prediction tower and the price prediction tower, the increase in training time is negligible compared to the base figure. Likewise, the inference time barely grows, thanks to the plug-in design of the AIE whose two auxiliary modules do not take effect at the inference phase.

\begin{table}[h]
	\centering
	\caption{Training time and inference time (whole test set) comparison on industrial dataset} 
	\vspace{-0.6em}
	\label{tab:eff}
	\resizebox{\linewidth}{!}{
	\begin{tabular}{c|c|ccccc}
		\toprule
		 Metric & Model & DCN & EDCN & DFFM & HierRec &  FiBiNET  \\
		\midrule
		 Training Time & Baseline &10.5& 11 & 11.5 &11.5 &11\\
	   (GPU Hour) &+AIE &11 & 11 & 11.5 & 12 & 11   \\
	 \cmidrule{1-7}
		 \multicolumn{1}{c|}{\multirow{2}{*}{Inference Time (s)}} & Baseline &553 & 612 & 562 &655 &632\\
	   &+AIE &556 & 615 & 569 & 663 & 636   \\
		
		\bottomrule
	\end{tabular}
}
\vspace{-0.6em}
\end{table}

\section{Related Work}
Our proposed AIE framework utilizes posterior auction information to enhance the CTR prediction model's performance. Therefore we would provide a brief overview of the literature related to the following two aspects. The related work of the methods involved like multi-task and multi-scenario learning will be covered briefly.

\textbf{CTR prediction}
CTR prediction models learn the user and item's relevance and hold a crucial place in recommender systems. Due to the importance of feature co-occurrence relation, feature interaction is vital to perform accurate CTR prediction. 
DeepFM \cite{guo2017deepfm} uses a FM component and places it parallel with the DNN to model feature interactions.
Deep \& Cross Network (DCN)~ \cite{wang2017deep} captures different orders' feature interaction by deploying layer-wise feature crossing recursively and DCN V2~\cite{wang2021dcn} upgrades the feature crossing vector to a matrix for enhancing representing ability.
EDCN \cite{chen2021enhancing} further enhances the performance by facilitating the information sharing between the parallel structures.
FiBiNET \cite{huang2019FiBiNET} combines the attention mechanism based on SENET \cite{hu2018squeeze} with a bi-linear interaction layer to dynamically learn feature weights and achieves fine-grained feature interactions. 
AutoInt~\cite{song2019autoint} achieves superior performance and good interpretability with a self-attention architecture to learn feature interactions.

Besides, multi-scenario recommendations~\cite{jiang2022adaptive,zang2022survey} are widely adopted to depict the differences in data distribution among different scenarios. To solve this problem, the Multi-Task Learning (MTL) paradigm~\cite{sharebottom} is widely used by constructing the shared and specific experts and applying adaptive gates to select relevant information for prediction.
Shared Bottom~\cite{sharebottom}, MMOE~\cite{mmoe} and PLE~ \cite{PLE} are representative models in MTL.
To achieve better modeling of multi-scenario, Dynamic Weight models~\cite{zhang2022leaving, yang2022adasparse} are proposed by generating scenario-specific dynamic parameters adaptively. 
DFFM~\cite{guo2023dffm} incorporates scenario-related information into the parameters of the feature interaction and user behavior modules, allowing for fine-grained scenario-specific learning.
Moreover, HierRec~\cite{gao2023scenario} conducts explicit and implicit scenario modeling simultaneously by a scenario-aware hierarchical dynamic network. 

\textbf{Auction Information Utilization}
As elaborated above, the CTR prediction problem is comprehensively optimized from the aspects of feature interaction, multi-task and multi-scenario. However, considering posterior auction information to enhance the CTR prediction is crucial in online advertising. MTAE~\cite{yang2021multi} proposed a framework to leverage posterior market price to ancillary CTR prediction. In the field of utility optimization, there is some related literature~\cite{wu2018turning, vasile2017cost, lyu2023pairwise} that utilizes auction information to optimize revenue. Although market price is considered for CTR prediction in online advertising, the exploitation of other auction information including bid is not enough as it can also provide extra posterior signals and impact the final online display exposure. Moreover, the fine-grained modeling of the market price is necessary to capture the distribution variance of the market price. Based on these considerations, we propose AIE to better utilize posterior auction signals to enhance the CTR prediction's performance.

\section{Conclusion}
In this paper, we delve into the problem of insufficient utilization of auction signals and first reveal the auction bias for CTR prediction in online advertising. Besides, a novel framework called AIE is proposed to better utilize the posterior auction information, which is lightweight, model-agnostic and latency-friendly. Specifically, AM2 constructs an auxiliary task to learn extra market price information while realizing a fine-grained perception of market prices in different auction scenarios. BCM performs a delicate reweighting method by using posterior bidding information to ease the auction bias and improve the prediction accuracy. Offline experiments are conducted on a public dataset and an industrial dataset to demonstrate its effectiveness and compatibility.
Besides, a one-month online A/B test in a large-scale advertising platform shows that AIE improves the base model by 5.76\% and 2.44\% in terms of eCPM and CTR. 

\bibliographystyle{ACM-Reference-Format}
\bibliography{acmart}

\end{document}